\documentclass[twocolumn,prb,aps,showpacs,footinbib,amsmath,amssymb,superscriptaddress,longbibliography,citeautoscript]{revtex4-2}
\usepackage{graphicx}
\usepackage{siunitx}  
\usepackage{comment}
\usepackage{booktabs}
\usepackage{dcolumn}
\usepackage{bm}
\usepackage{titlesec}
\usepackage{braket}
\usepackage[dvipsnames]{xcolor}
\usepackage[normalem]{ulem}

\newcommand{\1}{\c{c}}

\begin{document}
\title{
Implementation and application of a DFT$+U$$+V$ approach within the all-electron FLAPW method}
\author{W. Beida}
\email{w.beida@fz-juelich.de}
\affiliation{Peter Grünberg Institut, Forschungszentrum Jülich and JARA, 52425 Jülich Germany}
\affiliation{Institute for Theoretical Physics, RWTH Aachen University, 52062 Aachen, Germany}
\author{G. Bihlmayer}
\affiliation{Peter Grünberg Institut, Forschungszentrum Jülich and JARA, 52425 Jülich Germany}
\author{C. Friedrich}
\affiliation{Peter Grünberg Institut, Forschungszentrum Jülich and JARA, 52425 Jülich Germany}
\author{G. Michalicek}
\affiliation{Peter Grünberg Institut, Forschungszentrum Jülich and JARA, 52425 Jülich Germany}
\author{D. Wortmann}
\affiliation{Peter Grünberg Institut, Forschungszentrum Jülich and JARA, 52425 Jülich Germany}
\author{S. Blügel}
\affiliation{Peter Grünberg Institut, Forschungszentrum Jülich and JARA, 52425 Jülich Germany}
\affiliation{Institute for Theoretical Physics, RWTH Aachen University, 52062 Aachen, Germany}
\date{\today}

\begin{abstract}
We present an implementation of the density-functional theory DFT$+U$$+V$ formalism within the all-electron full-potential linearized augmented-plane-wave (FLAPW) method as implemented in the FLEUR code. The DFT$+U$$+V$ formalism extends DFT, supplemented by the onsite Coulomb interaction $U$, to address local correlation effects in localized states by incorporating intersite Coulomb interaction terms $V$. It holds promise for improving charge and bond disproportionation, charge and orbital ordering, charge density wave formation, charge transfer, and the intersite correlation resulting from  hybridization between states of neighboring sites in a solid. 
$U$ and $V$ parameters are obtained  from first principles using the constrained random-phase approximation (cRPA) employing two different atom basis representations to project the screened Coulomb interaction: the Wannier and the muffin-tin basis functions. We investigate in detail the impact of the $V$ term for typical  covalently bonded materials like graphene, for bulk semiconductors such as silicon and germanium, and for charge-transfer insulators like NiO. Our results demonstrate an improvement in accuracy of specific properties across these systems, providing a framework for describing materials with different interaction regimes. 
We compare our DFT$+U$$+V$ results using our cRPA parameter sets with (i) previous DFT$+U$$+V$  calculation employing pseudopotential approximations, (ii) with experimental results and (iii) with our $GW$ results. 
\end{abstract}
\maketitle
\section{\label{sec:level1}Introduction}
Despite the widespread success of density functional theory (DFT) in modeling the electronic structure of materials, its standard approximations often struggle to describe systems with strong electronic correlations. To address this limitation, the DFT$+U$ approach introduces a Hubbard correction that improves the treatment of localized $d$- and $f$-electron states~\cite{hubbard1964electron,shick1999implementation}. While the DFT$+U$ approach has proven to be valuable in capturing onsite correlation effects, it remains a single-site correction and neglects intersite Coulomb interactions. These non-local interactions can be essential for describing charge redistribution, orbital hybridization, and collective phenomena in correlated systems. The DFT$+U$$+V$ method~\cite{campo2010extended} promises to overcome these shortcomings by  supplementing the local with non-local interactions, enabling a more realiable treatment of electronically challenging systems. 

The significance of including intersite interactions becomes particularly beneficial across a variety of challenging material classes. For instance, in rare-earth nickelates such as NdNiO$_3$, DFT$+U$ fails to reproduce the experimentally observed metal-insulator transition and charge disproportionation~\cite{subedi2015low}. In charge-ordered systems like LuFe$_2$O$_4$, DFT$+U$ is able to stabilize the insulating charge-ordered states, but fails to select the experimentally observed ground state~\cite{xiang2007charge}. In actinide compounds such as UO$_2$, discrepancies remain in predicting the insulating gap and orbital ordering~\cite{freyss2012first}. In systems that exhibit more itinerant behavior, such as VO$_2$, DFT$+U$ struggles to capture accurate magnetic and structural properties due to the neglect of non-local interactions~\cite{stahl2020critical,haas2024incorporating}. 

In a seminal work, Wehling \textit{et al.}~\cite{wehling2011s} have shown that intersite Coulomb interactions play a decisive role in graphene, a covalently bonded two-dimensional (2D) system, for which  reduced screening, spatial confinement and covalency enhance electron-electron interactions and non-local correlation effects, influencing charge transfer and low-energy excitations. Thus, DFT$+U$$+V$ plays a crucial role in determining the electronic structure and emergent phenomena in 2D systems in general, \textit{e.g.}\ transition-metal chalcogenides and halides~\cite{haddadi2024site}. The inclusion of intersite Coulomb interactions allows capturing spatial charge fluctuations and non-local effects, which are key to understanding phenomena like charge density waves and unconventional magnetism~\cite{huang2020first, campetella2023electron}. For instance, in van der Waals 1T-TaS$_2$ monolayers, non-local Coulomb interactions renormalize the low-energy spectrum and modify Heisenberg superexchange interactions, offering a pathway for Coulomb engineering of magnetic ground states in 2D materials~\cite{chen2022controlling}. In Cr-based halides like CrI$_3$, non-local interactions influence magnetic anisotropy and interlayer coupling, which are essential for realizing 2D ferromagnetism~\cite{kim2019exploitable}. 

Building on these motivations, the DFT$+U$$+V$ method has been successfully applied to a wide range of correlated systems. It has improved the theoretical description of Mott insulators as MnO and NiO. In such systems, DFT$+U$$+V$ does not only improve the electronic structure (band gap and magnetic moment), but also the total energy and structural properties~\cite{campo2010extended,tancogne2020parameter}. It describes the charge ordering phenomena and stabilizes the correct ground state as in Fe$_{3}$O$_{4}$~\cite{anisimov1996charge}. 
An essential aspect emphasized in recent work by Marzari and co-workers~\cite{timrov2022accurate} is that the inclusion of $V$ extends the applicability of the functional beyond purely Mott–Hubbard systems to materials where charge-transfer processes dominate the low-energy physics. In such cases, the fundamental gap and electronic structure are determined not only by $U$ acting on transition-metal $d$-states, but also by the charge-transfer energy between metal and ligand orbitals. The DFT$+U$$+V$ approach with the non-local Coulomb interaction $V$ naturally incorporates this physics by energetically evaluating a redistribution of electronic charge between neighboring sites, thereby accounting for ligand-to-metal and metal-to-ligand hybridization effects that are essential for describing charge-transfer insulators and mixed-valence compounds. All these examples emphasize the need for an extended framework that can accurately account for both onsite and intersite electronic correlations in 2D and 3D systems.

In this work, we present the implementation of the DFT$+U$$+V$ functional within the full-potential linearized augmented plane-wave (FLAPW) framework, as realized in the FLEUR code~\cite{fleurWeb,fleurCode}. The FLAPW method is one of the most precise realizations of DFT for crystalline solids and thin films with unit cell compositions involving arbitrary elements of the periodic table~\cite{NatRevPhys-6-45-2024}. The implementation enables the accurate treatment of intersite Coulomb interactions in systems with hybridized bonding and strong correlation effects. The interaction parameters are determined from first principles using the constrained random phase approximation (cRPA) \cite{aryasetiawan2004frequency,sasioglu2011} as implemented in the SPEX code~\cite{SPEX2}. 

The values of $U$ and $V$ are known to depend sensitively on the definition of the correlated subspace and are therefore not universal nor directly transferable across different electronic-structure schemes. In this work, we examine two atom-centered localized representations for projecting the screened Coulomb interaction: (i) muffin-tin functions (MTFs), which ensure consistency within the LAPW framework, and (ii) maximally localized Wannier functions (MLWFs), which provide a natural bridge to low-energy model Hamiltonians. We compare the merits of these two choices and assess how the projector representation influences the resulting interaction parameters and the corresponding electronic properties.
\\

We validate our methodology by applying it to a diverse set of representative systems that span different bonding and correlation regimes. We begin with graphene, a prototypical two-dimensional material characterized by covalent bonding and highly delocalized $\pi$ electrons. Next, we examine the elemental semiconductors silicon and germanium, where hybridization provides a rigorous test for capturing intersite effects. Finally, we consider NiO, a well-known Mott-charge-transfer insulator, making it an ideal benchmark for strong electron-electron interactions. 

The remainder of this paper is organized as follows: In Sec.~\ref{sec:level2}, we introduce the extended DFT$+U$$+V$ formalism in the LAPW basis. The methodology for computing interaction parameters $U$ and $V$ via cRPA is discussed in Sec.~\ref{sec:level4}. The benchmark results and validation of the implementation are presented in Sec.~\ref{sec:level5}. We conclude in Sec.~\ref{sec:level6} with a summary and outlook.

\section{\label{sec:level2}Implementation of DFT$+U$$+V$}
In this section, we outline the theoretical and computational framework used to implement the DFT$+U$$+V$ approach. We begin with we briefly  introducing the necessary details and definitions  of the FLAPW method,  review the DFT$+U$ formalism. Then we extend the discussion to include the \mbox{DFT$+U$$+V$} method.

\subsection{The FLAPW method}
In general, the Kohn-Sham (KS) electron wave functions in periodic lattices have the form of Bloch waves, characterized by a Bloch vector, $\textbf{k}$, within the first Brillouin zone (BZ) of the reciprocal lattice, and a band index, $\nu$. In the LAPW method~\cite{Koelling_1975}, 
the KS wave function,  $\Psi_{\textbf{k},\nu}^{\sigma}(\textbf{r})$, is expanded into a set of basis functions, $\varphi^{\sigma}_{\textbf{k+G}}(\textbf{r})$ using the crystal potential, addressed by the reciprocal lattice vectors $\textbf{G}$: 
\begin{equation}
\begin{aligned}
\Psi_{\textbf{k},\nu}^{\sigma}(\textbf{r}) =
\sum_{\textbf{G}} c_{\textbf{k+G},\nu}^{\sigma} \varphi^{\sigma}_{\textbf{k+G}}(\textbf{r}), \quad \forall  \; |\textbf{k+G}| \leq K_{\text{max}}
\label{eq:eqn1neg}
\end{aligned}
\end{equation}
where $K_{\text{max}}$ is the main cutoff parameter of the LAPW basis controlling the accuracy of the calculation, and $c_{\textbf{k+G},\nu}^{\sigma}$ are the expansion coefficients. The FLAPW approach~\cite{wimmer1981flapw} is built upon a spatial partitioning into muffin-tin (MT) spheres, centered around atomic positions, and the interstitial region (IR) between these spheres. The wave functions, electron density, and potential are represented in a manner that respects this subdivision. 
The accuracy of the charge density and potential expansion is controlled by cut-off of their plane-wave expansions $G_\text{max}$.

 The LAPW basis functions are written as,
\begin{equation}
\begin{aligned}
\varphi_{\textbf{k+G}}^{\sigma}(\textbf{r}) =
\begin{cases}
\frac{1}{\sqrt{\Omega}} e^{i(\textbf{k+G}) \cdot \textbf{r}} &  \textbf{r}\in\text{IR} \\
\\
\sum_{L} \left(a^{I, L\sigma}_{\mathbf{k+G}} u^{\sigma}_{\ell} (r_{I}) +\right.  \\
\\ \left.\hspace{0.8 cm}b^{I,L\sigma}_{\mathbf{k+G}} \dot{u}^{\sigma}_{\ell}(r_{I})\right) Y_{L}(\hat{\mathbf{r}}_{I}) & \textbf{r}\in\text{MT}^{I},
\\
\end{cases}
\label{eq:eq0}
\end{aligned}
\end{equation}
where $\Omega$ is the volume of the unit cell. Within the MT sphere of atom $I$ and radius $R_\text{MT}$, the basis functions exhibit a dependence on $\mathbf{r}_{I}=\mathbf{r}- \mathbf{R}_{I}$, which represents the position vector relative to the atomic position $\mathbf{R}_{I}$. The wave functions are expanded using products of radial functions $\{u_{\ell},\dot{u}_{\ell}\}$ and spherical harmonics $Y_L(\hat{\mathbf{r}})$  with the unit vector $\hat{\mathbf{r}}=\mathbf{r}/r$. $L=({\ell},m)$ is a composite index comprising the angular momentum and magnetic quantum numbers.

The radial functions $u$ emerge as solutions to the scalar-relativistic approximation of the radial Dirac equation for a spherically averaged MT potential and an angular momentum and spin dependent energy parameter, and $\dot{u}$ are their energy derivatives. In the interstitial region, the wave functions are described by plane waves. The two representations are smoothly matched in both value and slope at the boundaries by matching coefficients $a$ and $b$, ensuring an accurate description of the electronic structure across the entire unit cell.

The expansion coefficients $c_{\textbf{k+G},\nu}^{\sigma}$ in Eq.~\eqref{eq:eqn1neg} can be contracted  with the matching coefficients $a$ and $b$ in Eq.~\eqref{eq:eq0} via the reciprocal lattice vectors $\textbf{G}$ to facilitate the simplification of the spin density and the occupation matrix representations in the next sections by defining the matching coefficients for the Kohn-Sham eigenfunctions at atom $I$: $A^{I,L\sigma}_{\textbf{k},\nu}=\sum_{\textbf{G}} c_{\textbf{k+G},\nu}^{\sigma} a^{I,L\sigma}_{\textbf{k+G}}$, and $B^{I,L\sigma}_{\textbf{k},\nu}=\sum_{\textbf{G}} c_{\textbf{k+G},\nu}^{\sigma} b^{I,L\sigma}_{\textbf{k+G}}$. 
The valence spin density in the muffin-tin sphere of atoms $I$ is then given by the expression
\begin{eqnarray}\label{eq:spin-density}
n^{I,\sigma}(\mathbf{r}) 
&=& \sum_{\nu} \sum_{\mathbf{k}}^{\text{BZ}} w_{\mathbf{k}} f^{\sigma}_{\mathbf{k},\nu} 
\sum_{L,L{'}}
\Big(\\&\phantom{+}&
A^{*I,L\sigma}_{\textbf{k},\nu} A^{I,L{'}\sigma}_{\textbf{k},\nu} u^\sigma_{\ell}(r_I) u^\sigma_{\ell'}(r_I) \nonumber \\
&+& B^{*I,L\sigma}_{\textbf{k},\nu} A^{I,L{'}\sigma}_{\textbf{k},\nu} \dot{u}^\sigma_{\ell}(r_I) u^\sigma_{\ell'}(r_I) \nonumber \\
&+& A^{*I,L\sigma}_{\textbf{k},\nu} B^{I,L{'}\sigma}_{\textbf{k},\nu}  u^\sigma_{\ell}(r_I) \dot{u}^\sigma_{\ell'}(r_I)\nonumber \\
&+& B^{*I,L\sigma}_{\textbf{k},\nu} B^{I,L{'}\sigma}_{\textbf{k},\nu} \dot{u}^\sigma_{\ell}(r_I) \dot{u}^\sigma_{\ell'}(r_I)
\Big)
Y^{*}_{L}(\hat{\mathbf{r}}_I) Y_{L{'}}(\hat{\mathbf{r}}_I)\nonumber,
\end{eqnarray}
where  $w_{\mathbf{k}}$ is the $\mathbf{k}$-point weight for the integration over the BZ and the Dirac function  $f^{\sigma}_{\mathbf{k},\nu}$ separates occupied from non-occupied Bloch states $|\Psi^{\sigma}_{\mathbf{k},\nu}\rangle$ of band index $\nu$.

\subsection{DFT$+U$}
One widely used approach to improve the accuracy of DFT for strongly correlated systems or systems with localized electrons exhibiting a strong self-interaction is the DFT$+U$ method. This technique incorporates a Hubbard-like correction term to better describe the onsite Coulomb interactions among localized electrons. Thus, it corrects the tendency of standard DFT to overly delocalize these electrons, while preserving the conventional DFT treatment for itinerant, weakly correlated electrons~\cite{shick1999implementation, agapito2015reformulation,campo2010extended}. To achieve this, DFT$+U$ introduces a correction to the total-energy functional of the form:
\begin{equation}
\begin{aligned}
E_{\text{DFT}+U} [n^{\sigma},\{ n_{mm{'}}^{I, \ell\sigma} \}]&= E_{\text{DFT}}[n^{\sigma}] + E_{U}[\{ n_{mm{'}}^{I, \ell \sigma} \}] \\ & =E_{\text{DFT}}[n^{\sigma}]+ E_{\text{ee},U}[\{ n_{mm{'}}^{I,\ell \sigma} \}] \\&- E_{\text{dc},U}[\{ n_{mm{'}}^{I,\ell \sigma} \}],
  \label{eq:eq1}
\end{aligned}
\end{equation}
where $n^{\sigma}$ denotes the spin density. 
The term $E_{\text{ee},U}$ accounts for the electron-electron interaction energy computed via the Hartree-Fock approximation within the Hubbard model for correlated electrons on the same atom $I$, while $E_{\text{dc},U}$ is a double-counting correction that eliminates the interaction terms for correlated electrons already included in DFT. Both terms depend on the density matrix of localized orbitals $m, m{'}$ in subspace $\ell$   of atom $I$, which is to be corrected, denoted by 
\begin{equation}
\label{eq:orbital-occupation_onsite}
n^{I,\ell\sigma}_{mm'} = \sum_{\nu} \sum_{\mathbf{k}}^{\text{BZ}} w_{\mathbf{k}} f^{\sigma}_{\mathbf{k},\nu} 
\left\langle \Psi^{\sigma}_{\mathbf{k},\nu} \middle| \hat{P}^{I,\ell\sigma}_{mm'} \middle| \Psi^{\sigma}_{\mathbf{k},\nu} \right\rangle 
\end{equation}
where $\hat{P}^{I,\ell\sigma}_{mm'}=|\phi_{{\ell}m}^{I,\sigma} \rangle \langle\phi_{{\ell}m^{\prime}}^{I,\sigma} | $ with $m,m{'}\in\mathbb{Z}\cap\{-\ell, \dots , \ell\}$ is the projector associated with the localized orbitals  $|\phi_{\ell m}^{I,\sigma} \rangle$ that form the basis of the Hilbert space in which the electron correlation is corrected. $w_{\mathbf{k}}$ and the Dirac function  $f^{\sigma}_{\mathbf{k},\nu}$ are the typical weights defined in Eq.~\eqref{eq:spin-density}. 
Physically, $n^{I,\ell\sigma}_{mm'}$ gives the occupation matrix for the specified orbitals $m,m{'}$. Together, these terms define the correction $E_{U}$, which is added to the total energy of DFT. 

The implementation in the FLEUR code follows the rotationally invariant formulation proposed by Shick \textit{et al.}~\cite{shick1999implementation}.
The localized orbitals used to project the electronic states into the corrected Hilbert space are defined as the muffin-tin functions (MTFs) centered on atom $I$ and characterized by angular momentum $\ell$
\begin{equation}\label{eq:MTF}
\phi_{L}^{I,\sigma}(\mathbf{r})=
\left(u^{I,\sigma}_{{\ell}_{I}}({r}) + 
\frac{1}{\dot{N}^{I,\sigma}_{{\ell}_{I}}}\dot{u}^{I,\sigma}_{{\ell}_{I}}({r})\right)Y_{L_{I}}(\hat{\mathbf{r}})
\end{equation}
with $\phi_{L}^{I,\sigma}(\mathbf{r})=0$ for $r>R_\text{MT}$. The onsite occupation matrix is then given  by 
\begin{eqnarray}
\label{eq:orbital-occupation_onsite_FLAPW}
n^{I,\ell\sigma}_{mm'} = 
\sum_{\nu} \sum_{\mathbf{k}}^{\text{BZ}} w_{\mathbf{k}} f^{\sigma}_{\mathbf{k},\nu} \!\!&\Big(& \!\!
    A_{\mathbf{k}, \nu}^{*I,\ell m{'} \sigma} A_{\mathbf{k}, \nu}^{I,\ell m\sigma} \\
    &+& B_{\mathbf{k}, \nu}^{*I,\ell m{'} \sigma} B_{\mathbf{k}, \nu}^{I,\ell m\sigma} 
        (\dot{N}^{I,\sigma}_{{\ell}_{I}})^{2}\Big)\, , \nonumber
\end{eqnarray} 
where  $(\dot{N}^{I,\sigma}_{{\ell}_{I}})^2= \langle \dot{u}^{I,\sigma}_{{\ell}_{I}}| \dot{u}^{I,\sigma}_{{\ell}_{I}}\rangle$ denotes  the normalization constant.

\subsection{DFT$+U$$+V$}
To account for intersite electronic interactions a generalized occupation matrix was introduced in Ref.~\cite{campo2010extended} . The Kohn-Sham wave functions are projected  onto atomic basis functions associated with two distinct atoms, $I$ and $J$, $I\ne J$,
\begin{equation}
    \begin{aligned}
n_{L_{I} L_{J}}^{IJ,\sigma} &=\sum_{\mathbf{k},\nu} w_{\mathbf{k}}f^{\sigma}_{\mathbf{k},\nu} \langle \Psi_{\mathbf{k},\nu}^{\sigma}|\phi_{L_{J}}^{J,\sigma} \rangle  \langle \phi_{L_{I}}^{I,\sigma}|\Psi_{\mathbf{k},\nu}^{\sigma}\rangle .
    \end{aligned}
    \label{eq:eq6}
\end{equation}
Here, $L_{I}$ and $L_{J}$ denote ${\ell}_{I}, {\ell}_{J}$ and their corresponding $m_{I}, m_{J}$ orbitals to which we apply the intersite Hubbard correction $V$. Setting $I = J$ in Eq.~\eqref{eq:eq6} recovers the onsite occupation matrix of Eq.~\eqref{eq:orbital-occupation_onsite}. In the FLAPW implementation, the MTF formalism Eq.~\eqref{eq:MTF} is extended such that for each atomic site and spin channel,  the  local basis is spanned by $\{ u_{\ell}, \dot{u}_{\ell}/\dot{N}^{\sigma}_{\ell}\}$ multiplied by  spherical harmonics $Y_{L}$. Consequently, the projector $P^{IJ,\sigma}_{L_{I} L_{J}}$  between atoms $I$ and $J$ and the associated orbitals $L_{I}$ and $L_{J}$ is expressed as follows: 
\begin{eqnarray}
&&|\phi_{L_{J}}^{J,\sigma} \rangle 
 \langle \phi_{L_{I}}^{I,\sigma}|= |u^{J,\sigma}_{{\ell}_{J}}Y_{L_{J}} \rangle 
 \langle u^{I,\sigma}_{{\ell}_{I}}Y_{L_{I}}|+ 
\frac{|u^{J,\sigma}_{{\ell}_{J}}Y_{L_{J}}\rangle
 \langle \dot{u}^{I,\sigma}_{{\ell}_{I}}Y_{L_{I}}|}{\dot{N}^{I,\sigma}_{{\ell}_{I}}}\nonumber \\  && + \frac{|\dot{u}^{J,\sigma}_{{\ell}_{J}}Y_{L_{J}}\rangle 
 \langle u^{I,\sigma}_{{\ell}_{I}}Y_{L_{I}}|}{\dot{N}^{J,\sigma}_{{\ell}_{J}}} +
\frac{|\dot{u}^{J,\sigma}_{{\ell}_{J}}Y_{L_{J}} \rangle 
 \langle \dot{u}^{I,\sigma}_{{\ell}_{I}}Y_{L_{I}}|}{\dot{N}^{I,\sigma}_{{\ell}_{I}} \dot{N}^{J,\sigma}_{{\ell}_{J}}}\, .
\label{eq:eq7}
\end{eqnarray}
Thus, the intersite occupation matrix is expressed by
\begin{align}
n_{L_I L_J}^{I J, \sigma} = \sum_{\mathbf{k}, \nu} & w_{\mathbf{k}}f_{\mathbf{k}, \nu}^{\sigma} \Big(
    A_{\mathbf{k}, \nu}^{*J,L_J \sigma} A_{\mathbf{k}, \nu}^{I,L_I \sigma} \notag \\
    &+ A_{\mathbf{k}, \nu}^{*J,L_J \sigma} B_{\mathbf{k}, \nu}^{I,L_I \sigma} 
        \dot{N}^{I,\sigma}_{{\ell}_{I}} \notag \\
    &+ B_{\mathbf{k}, \nu}^{*J,L_J \sigma} A_{\mathbf{k}, \nu}^{I,L_I \sigma} 
        \dot{N}^{J,\sigma}_{{\ell}_{J}} \notag \\
    &+ B_{\mathbf{k}, \nu}^{*J,L_J \sigma} B_{\mathbf{k}, \nu}^{I,L_I \sigma} 
        \dot{N}^{I,\sigma}_{{\ell}_{I}} \dot{N}^{J,\sigma}_{{\ell}_{J}} \Big)
        \label{eq:eq8}
\end{align}
 The matrix $n^{IJ,\sigma}_{L_{I} L_{J}}$ is related to $n^{JI,\sigma}_{L_{J} L_{I}}$ by Hermitian conjugation. In the present implementation, the spin-projection is not site dependent ($\sigma_I=\sigma_J$), \text{i.e.}\ it is limited to collinear spin configurations and the absence of spin-orbit interaction, and subsequently a single spin index is used for the intersite occupation matrix. While different choices of projectors are possible~\cite{haule2010dynamical}, in the present choice, the $AB$ and $BA$ terms remain, unlike in the onsite projector Eq.~\eqref{eq:orbital-occupation_onsite}. These mixed terms are important in scenarios where one atom contributes primarily through $A$, while the other contributes only through $B$. In such cases, the $AA$ and $BB$ terms vanish, making the $AB$ and $BA$ contributions essential. These mixed contributions do not appear in the on-site case. 

It is worth mentioning that local orbitals (LOs) are not included in our projector construction. LOs are supplementary basis functions confined within the muffin-tin spheres, typically used to accurately describe semi-core states~\cite{singh1991ground}. However, since semi-core states are not part of our correlated subspace, excluding LOs allows us to maintain a simpler formalism without sacrificing the accuracy of our results.
We adopt the same form of energy correction, including the double counting term, as proposed by Campo \textit{et al.}~\cite{campo2010extended}, given by 
\begin{equation}
    E_{V}=E_{\text{ee},V} - E_{\text{dc}, V} =  -\sum^{*}_{IJ} \frac{V^{IJ}}{2}  \sum_{\sigma} Tr(n^{IJ,\sigma} n^{JI,\sigma})\, ,
    \label{eq:equation_Venergy}
\end{equation}
where $V^{IJ}$ represents the interaction matrix between the orbitals of atomic sites $I$ and $J$. The exact form of the interaction matrix $V^{IJ}$ is discussed in Sec.~\ref{sec:level4}.  The sum with the star includes for each atom $I$ its neighboring atoms $J$ within a specified distance and chemistry. As in the DFT$+U$ method, there is no unique form for the double-counting correction $E_{\text{dc}}$; rather, approximations are typically employed. For the $U$ correction, $E_{\text{dc},U}$ is either approximated in the fully localized limit, or the around-mean-field limit~\cite{petukhov2003correlated,bultmark2009multipole}, where orbital occupations are treated as partially filled in an averaged manner. For the $V$ correction, the double counting correction is considered in the mean field limit~\cite{campo2010extended}.  Employing Eq.~\eqref{eq:equation_Venergy} and taking the functional derivative of the energy with respect to the Kohn-Sham wave function, one obtains the potential matrix that is responsible for introducing the intersite interaction correction, 
\begin{eqnarray}
\label{eq:eq9}
\hspace{-0.5cm}\langle \varphi_{\mathbf{k+G_{1}}}^{\sigma} |\hat{V}_{V}| \varphi_{\mathbf{k+G_{2}}}^{\sigma} \rangle 
&=& -\sum_{IJ}^{*} V^{IJ}
\sum_{L_I L_J}  
n_{L_J L_I}^{JI,\sigma}
\Big(                      \notag \\
&&\phantom{+} a^{*I,L_I \sigma}_\mathbf{k+G_{1}} a^{J,L_J \sigma}_\mathbf{k+G_{2}}  \notag \\
&& + b^{*I,L_I \sigma}_\mathbf{k+G_{1}} a^{J,L_J \sigma}_\mathbf{k+G_{2}} 
\dot{N}^{I,\sigma}_{{\ell}_{I}}  \notag \\
&& + a^{*I,L_I\sigma}_\mathbf{k+G_{1}} b^{J,L_J\sigma}_\mathbf{k+G_{2}}  
\dot{N}^{J,\sigma}_{{\ell}_J} \notag \\
&& + b^{* I,L_I \sigma}_\mathbf{k+G_{1}} b^{J,L_J \sigma}_\mathbf{k+G_{2}} 
\dot{N}^{I,\sigma}_{{\ell}_I} \dot{N}^{J,\sigma}_{{\ell}_J}
\Big)\, .
\end{eqnarray}
 This represents the correction term to the complete intersite electronic Hamiltonian. The next step is to compute the $V^{IJ}$ parameters, which we will determine using cRPA.

\section{\label{sec:level4}cRPA approach for Hubbard parameters calculation}
Constrained random-phase approximation (cRPA) is a method for calculating the screened Coulomb interaction between electrons. It enables the determination of the full Coulomb matrix, from which one can extract the onsite, exchange, and intersite interactions, both in the static and frequency-dependent regimes~\cite{aryasetiawan2004frequency,aryasetiawan2011constrained}. The fully screened Coulomb interaction $W$ is related to the bare Coulomb interaction $V_{\text{Bare}}$ by
\begin{equation}
    \begin{aligned}
    W(\mathbf{r},\mathbf{r}{'}; \omega)= \int d\mathbf{r}{''} \epsilon^{-1}(\mathbf{r},\mathbf{r}{''}; \omega) V_{\text{Bare}}(\mathbf{r}{''},\mathbf{r}{'})\, ,
    \label{eq:eq10}
    \end{aligned}
\end{equation}
where $\epsilon (\mathbf{r},\mathbf{r}{''}; \omega)$ is the dielectric function. It is connected to the electron polarization function $P$ through the relationship
\begin{equation}
    \begin{aligned}
    \epsilon(\mathbf{r},\mathbf{r}{'}; \omega)= \delta(\mathbf{r}-\mathbf{r}{'}) - \int d\mathbf{r}{''} V_{\text{Bare}}(\mathbf{r},\mathbf{r}{''}) P(\mathbf{r}{''},\mathbf{r}{'};\omega)\, .
    \label{eq:eq11}
    \end{aligned}
\end{equation}
The electron polarization function is given by the expression
\begin{equation}
    \begin{aligned}
    P(\mathbf{r},\mathbf{r}{'};\omega)=& 2 \sum_{\mu}^{\text{occ}} \sum_{\nu}^{\text{unocc}} \Psi_{\mathbf{k},\mu}(\mathbf{r}) \Psi^{*}_{\mathbf{k}',\nu}(\mathbf{r}) \Psi^{*}_{\mathbf{k},\mu}(\mathbf{r}{'}) \Psi_{\mathbf{k}',\nu}(\mathbf{r}{'}) \\&
    \left( \frac{1}{\omega-\Delta_{\mathbf{k},\mu;\mathbf{k}',\nu}+i \eta} - \frac{1}{\omega+\Delta_{\mathbf{k},\mu;\mathbf{k}',\nu}-i \eta} \right) ,
    \label{eq:eq12}
    \end{aligned}
\end{equation}
where $\eta$ is a positive infinitesimal, and $\Delta_{\mathbf{k},\mu;\mathbf{k}',\nu}= E_{\mathbf{k}',\nu}- E_{\mathbf{k},\mu}$ is the difference between Kohn-Sham eigenvalues. The Coulomb matrix is calculated in the non-magnetic phase of the materials. Therefore, there is no explicit spin index, and a spin factor of 2 is included. Consequently, the spin index is omitted in this section. In the cRPA method, to exclude the screening effects arising from the correlated subspace $l_{\text{c}}$ the total polarization function is split into two components, $P= P_{l_{\text{c}}}+P_{r}$, where $P_{l_{\text{c}}}$ accounts for transitions between states within the correlated subspace, and $P_{r}$ describes transitions outside of this subspace~\cite{sasioglu2011}. The relationship between the screened Coulomb interaction, polarization, and bare interaction is given by the matrix equation,
\begin{equation}
    \begin{aligned}
    U(\omega)= [1-V_{\text{Bare}}P_{r}(\omega) ]^{-1} V_{\text{Bare}}\, .
    \label{eq:eq13}
    \end{aligned}
\end{equation}

In the last step, the partially screened Coulomb interaction $U(\omega)$ is represented in an atom-centered localized basis.
We consider two such basis sets in this work, the maximally localized Wannier functions (\mbox{MLWFs})~\cite{wannier} $\{w_{L\mathbf{R}}(\mathbf{r})\}$ and a set based on the MT functions (MTF) of the LAPW basis, $\{u_{L\mathbf{R}}(\mathbf{r})\}$. The vector $\mathbf{R}$ is the corresponding position vector of the atomic site and $L$ is the orbital index. The MTF $u_{L\mathbf{R}}(\mathbf{r})=u_{\ell}(|\mathbf{r-R}|)Y_{L}(\widehat{\mathbf{r-R}})$, with the radial functions $u_{\ell}(r)$ and the spherical harmonics $Y_{L}(\hat{\mathbf{r}})$ are non-zero only in the MT sphere of the atom at $\mathbf{R}$, \textit{i.e.}\ with $u_{\ell}(r)=0$ if $r$ is larger than the MT radius of that atom. Just as the MLWFs, the functions $u_{L\mathbf{R}}(\mathbf{r})$ are orthonormal and are a good approximation to the space spanned by the projector Eq.~\eqref{eq:eq8}. The second set has been included for comparison, as it is consistent with the DFT$+U$$+V$ implementation, which uses the same atom-centered basis functions. For the sake of consistency, the same correlated subspace is used, in particular the same polarization function $P_{r}(\omega)$ in Eq.~\eqref{eq:eq13}, for the two localized basis sets.

For the MLWFs, the m-matrix representation of the screened Coulomb interaction can be expressed for each $\mathbf{R}$ as
\begin{equation}
    \begin{aligned}
    U_{\mathbf{R}m_{1},m_{2},m_{3},m_{4}}(\omega)=\int d\mathbf{r} d\mathbf{r}{'} & \,w^{*}_{m_{1}\mathbf{0}}(\mathbf{r}) w_{m_{3}\mathbf{0}}(\mathbf{r}) \\& U(\mathbf{r},\mathbf{r}{'}; \omega) \\& w^{*}_{m_{2}\mathbf{R}}(\mathbf{r}{'}) w_{m_{4}\mathbf{R}}(\mathbf{r}{'})\, .
    \label{eq:eq14}
    \end{aligned}
\end{equation}
We use an analogous definition of the m-matrix representation of the screened Coulomb interaction for the set of MTF $\{u_{L\mathbf{R}}\}$, but we should keep in mind that the values of the matrix elements in these different representations can be very different. In the static limit and $\mathbf{R}=\mathbf{0}$, the orbital m-averaged onsite diagonal interaction $U$, exchange interaction $J$, and intersite interaction $V$ of the screened Coulomb potential are defined as:
\begin{equation}
    \begin{aligned}
    U_{{\ell}_{I}}= \frac{1}{D_{\ell_I}^{2}} \sum_{m,m{'}=-{\ell}_{I}}^{{\ell}_{I}} U^{II}_{mm{'},mm{'}}(\omega=0)\, ,
    \label{eq:eq15}
    \end{aligned}
\end{equation}
\begin{equation}
    \begin{aligned}
    J_{{\ell}_{I}}=U_{{\ell}_{I}}- \frac{1}{D_{\ell_I}(D_{\ell_I}-1)} \sum_{m,m{'}=-{\ell}_{I}}^{{\ell}_{I}}\!\!\!\!\! & \left(U^{II}_{mm{'},mm{'}}-\right.\\&\left.U^{II}_{mm{'},m{'}m}\right)(\omega=0)\, ,
    \label{eq:eq16}
    \end{aligned}
\end{equation}
\begin{equation}
    \begin{aligned}
    V_{{\ell}_{I} {\ell}_{J}}= \frac{1}{D_{\ell_I} D_{\ell_J}}  \sum_{m=-{\ell}_{I}}^{{\ell}_{I}} \sum_{m{'}=-{\ell}_{J}}^{{\ell}_{J}} U^{IJ}_{mm{'},mm{'}}(\omega=0),
    \label{eq:eq17}
    \end{aligned}
\end{equation}
where $D_{\ell_{I(J)}}= 2\ell_{I(J)} +1$ is the degeneracy of the $\ell_{I(J)}$ level. We have added the superscripts $II$ and $IJ$ on the right-hand side of Eq.~\eqref{eq:eq15}–Eq.~\eqref{eq:eq17} to clarify that the indices $m$ and $m'$ refer to atom $I$ in Eq.~\eqref{eq:eq15} and Eq.~\eqref{eq:eq16}, and to atoms $I$ and $J$ in Eq.~\eqref{eq:eq17}. The Hubbard matrix reveals that $U_{{\ell}_{I}}$ corresponds to the m-average of the diagonal block related to  atom $I$, while $V_{{\ell}_{I} {\ell}_{J}}$ represents the average of the off-diagonal block between atom $I$ and $J$. 
A brief note: In the results sections, we denote the parameter sets obtained using Wannier functions as $w$-cRPA, and those obtained using MTF functions as $u$-cRPA, but both are always applied to projectors spanned by MTF functions. 

In the $w$-cRPA approach, the screened Coulomb interaction is projected on MLWFs that are constructed using the \textsc{Wannier90} library~\cite{Mostofi2014} employed by the SPEX code. The MLWFs are expressed as linear combinations of Bloch states expanded in the LAPW basis, and thus include both the radial functions inside the MT spheres and IR contributions. In the $u$-cRPA calculations we employ the localized MTF-functions $\{u\}$ on which the Coulomb interaction is projected,  which are MT-radii dependent and are constructed directly from the self-consistent potential obtained from DFT calculations performed with the FLEUR code. By increasing the radii of the muffin-tin spheres, the $u$-functions acquire more weights. As a result, the $u$-cRPA interaction parameters become closer in value to those obtained from the $w$-cRPA approach, similar to trends observed in pseudopotential-based implementations~\cite{tancogne2020parameter}.
\section{\label{sec:level5}Computational details, results and discussion}
With the formulation of DFT$+U$$+V$ established in the LAPW basis set and the Hubbard parameters calculated, we are now prepared to benchmark our implementation and compare electronic, structural, and magnetic properties with published results and $GW$ calculations for a diverse spectrum of test systems: 2D Dirac semimetal graphene, covalently bonded semiconducting bulk Si and Ge, and insulating antiferromagnetic bulk NiO.
\\

For all systems, DFT calculations were performed using the generalized gradient approximation (GGA) for the exchange-correlation potential, as parameterized by Perdew, Burke, and Ernzerhof (PBE)~\cite{perdew1996generalized}. The equilibrium lattice constant and bulk modulus were obtained by analyzing the dependence of the ground-state energy on the lattice volume and fitting the resulting data to the Birch–Murnaghan equation of state~\cite{birch1947finite}. This we refer to in the results sections below, as DFT results.
The cRPA calculations were then performed at the DFT-relaxed equilibrium lattice parameters

\subsection{\label{sec:level5-1}Graphene}  
In a seminal study 
by Wehling \textit{et al.}~\cite{wehling2011s}, it was recognized that the non-local Coulomb interaction is unexpectedly large in the prototype 2D material graphene. The study presents Wannier-based cRPA calculations of both onsite and intersite interaction parameters up to the fourth nearest neighbors, focusing on the $p_z$-orbitals of carbon atoms. In a more recent work of Tancogne-Dejean and Rubio~\cite{tancogne2020parameter}, the onsite and intersite interaction parameters were calculated using a hybrid-like functional derived from the extended Hubbard model, applied to the carbon $p$-orbitals.

In this work, as a first step, we performed a DFT ground-state calculation using a $30 \times 30$ k-mesh. For the augmented plane-wave basis, a cutoff radius of $K_\text{max}=4.5$~$a_\text{B}^{-1}$ in reciprocal space was applied, along with  
$G_\text{max} = 13.5$~$a_\text{B}^{-1}$,
and an angular momentum cutoff of ${\ell}_\text{max}=6$ within the muffin-tin spheres for Carbon atoms. The MT radius of the Carbon atoms is $1.31$~$a_\text{B}$. As in previous studies~\cite{wehling2011s,tancogne2020parameter} the calculations are performed for graphene at its equilibrium lattice constant of $a_0=2.47$~\AA.

For later comparison, we also performed an one-shot many-body perturbation theory calculation within the $GW$ approximation utilizing the SPEX code~\cite{SPEX2}, following the methodology outlined in~\cite{friedrich2010efficient}. We employ a $36 \times 36$ k-point grid and the cutoff parameters for the mixed product basis were defined as angular momentum quantum number $L_{\textrm{cut}}=4$ and reciprocal cutoff $G_{\textrm{cut}}=4.5$~$a_\text{B}^{-1}$. Subsequently, we calculated the Hubbard parameters $U$ and $V$ for a $p$-model using a $16 \times 16$ grid, where the correlated subspace consists of the carbon $p$-orbitals in both Wannier and MTF bases.

\begin{table}[ht]
\caption{\label{table:table1} 
Partially screened onsite ($U$) and intersite ($V_{0i}$, up to third nearest neighbor) interaction parameters for graphene, calculated using cRPA using $u$- and $w$-functions. The results are compared with values from Refs.~\cite{wehling2011s, tancogne2020parameter}. The last column shows the parameters obtained by fitting the band structure to $GW$ results.}
\begin{ruledtabular}
\begin{tabular}{lccccc}
Parameter (eV) & $u$-cRPA & $w$-cRPA & Ref.~\cite{wehling2011s} & Ref.~\cite{tancogne2020parameter} & fit \\
\midrule
$U$      & 14.15 & 8.46 & 9.3  & 7.58 & 8.89 \\ 
$V_{01}$ & \phantom{1}6.39  & 5.26 & 5.5  & 4.00 & 3.20 \\ 
$V_{02}$ & \phantom{1}4.15  & 3.37 & 4.1  & 2.56 & 2.04 \\    
$V_{03}$ & \phantom{1}3.69  & 2.74 & 3.6  & 2.22 & 1.48 \\ 
\end{tabular}
\end{ruledtabular}
\end{table}

 The bare and partially screened $u$-cRPA and $w$-cRPA are collected in Table I and the Supplementary Table~II~\cite{SM}. We find that the $u$-cRPA parameters are larger than their Wannier counterparts. This discrepancy arises from the differences in orbital localization between the two basis sets. In graphene, the $p$-orbitals contribute to covalently bonding and antibonding $\pi$-bands, which are delocalized across the lattice or even localized between atoms, resulting in spatially extended Wannier functions. Conversely, the spatial extent of the $u$-functions is tightly constrained by the MT sphere around the carbon atoms, leading to greater localization. Our $w$-cRPA parameters show small deviations from those reported by Wehling \textit{et al.}~\cite{wehling2011s}, which may be due to the choice of the correlated subspace. 
 While in~\cite{wehling2011s} only the $p_{z}$-orbitals were included, our calculations employ a subspace composed of all $p$-orbitals. 
 As a consequence, the Wannier functions have a slightly different shape and spread, which affects the parameters. 
 Furthermore, small discrepancies in off-site parameters may also be caused by a change in the $\mathbf{k}$-integration routines. In newer versions of the code, the phase factor $\exp(i\mathbf{kR})$ originating from the off-site shift $\mathbf{R}$ is integrated analytically rather than being summed over. As said above the values or $U$ and $V$ parameters are not universal, but depend on the projector orbitals and thus on the context of use. The discrepancy between our parameters and those reported by Tancogne-Dejean and Rubio in~\cite{tancogne2020parameter} may be attributed to the different methods used for treating screening, different spatial spread of the projector orbitals in real space, as well as the fact that their parameters were computed self-consistently as part of the energy minimization process, whereas we calculated them in a single-step procedure.

\begin{figure}[h!]
\includegraphics[width=\columnwidth]{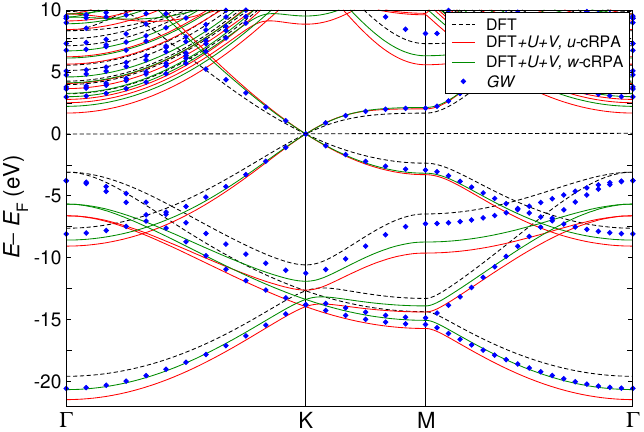}
\caption{\label{fig:dftUV_Graphene} Band structure of graphene calculated using GGA (dashed black lines) and GGA+$U$$+V$, using both $u$- and $w$- cRPA parameters (red and green lines). The results are compared to the $GW$ band structure (blue dots).}
\end{figure}

Using our calculated screened cRPA parameters in $u$- and $w$-function bases, we performed DFT$+U$$+V$ calculations and compared the resulting band structures with those from DFT and $GW$, presented in Fig.~\ref{fig:dftUV_Graphene}. At first we draw our attention to the Fermi velocity, $v_{\text{F}}$, as a critical measurable quantity governing graphene's electronic transport properties. It is calculated from the slope at the Dirac point located at the K point. As summarized in Table~\ref{table:table4}, standard DFT significantly underestimates $v_{\text{F}}$, while $GW$ provides a substantial improvement by incorporating many-body corrections to quasiparticle energies, thereby refining the slope of the Dirac cone. A modest improvement is observed when accounting for local correlations within DFT$+U$. Notably, our cRPA-based DFT$+U$$+V$ calculations exhibit strong agreement with experimental data, underscoring their effectiveness in capturing both local and nonlocal Coulomb interactions. The improved agreement between our results and the experimental Fermi velocity can be attributed to our larger interaction parameters in comparison to Ref.~\cite{tancogne2020parameter}. Considering the large differences of the $U$ and $V$ values when projecting the cRPA result on the  two  different projector functions $u$ and $w$, it is surprising that both Fermi velocities are so close and so close to the experimental data. It turns out $u$($w$)-cRPA slightly over-(under-)estimated the experimental value.

\begin{table}[ht]
\caption{\label{table:table4} 
Fermi velocity, $v_{\text{F}}$, of the $\pi^{*}$ band in graphene near the K point, calculated using various theoretical approaches. Results are compared with experimental data from Ref.~\cite{bostwick2007quasiparticle}.}
\begin{ruledtabular}
\begin{tabular}{lc}
Approach & $v_{\text{F}}$ ($\times 10^{5}$ m\,s$^{-1}$) \\
\midrule
DFT (this work / Ref.~\cite{tancogne2020parameter} ) & 8.32 / 7.04 \\ 
DFT+$U$ (fit)                                   & 8.56 \\ 
DFT$+U$$+V$ ($u$/$w$)-cRPA                        & 11.11 / 10.74 \\ 
DFT$+U$$+V$ (this work, fit / Ref.~\cite{tancogne2020parameter}) & 9.76 / 9.62 \\     
$GW$                                            & 10.27 \\ 
Experiment~\cite{bostwick2007quasiparticle}     & 11.00 \\
\end{tabular}
\end{ruledtabular}
\end{table}

Next, we focus in Fig.~\ref{fig:dftUV_Graphene} on the overall comparison of the results from DFT+$U$+$V$ with the results from DFT (GGA) and $GW$ for the occupied states over an energy range of 20~eV. Both the red ($u$-cRPA) and green ($w$-cRPA) curves reproduce the qualitative behavior of the $GW$ (blue dots) results well, capturing the overall dispersion and curvature of the bands. However, some quantitative differences are visible: Overall the $w$-cRPA results follows the $GW$ results a bit closer. The errors are largest for occupied states with a binding energy of about $5$~eV. This suggests that the static limit in the determination of the interaction parameters $U$, $J$ and $V$ (see Eq.~\eqref{eq:eq15}--Eq.~\eqref{eq:eq17}) might be a too strong of an approximation to improve and describe  systematically an energy scale of $20$~eV suggesting the application of and energy dependent $U$ or $V$ in the future.

We examine how far one can go with static parameters and fit the values of $U$ and $V$ directly to the $GW$ results  through an iterative parameter refinement approach using the $U$ and $V$ are derived from cRPA as initial values. At the end of this optimization process, the resulting parameter set, referred to as “fit”, represents the tuned $U$ and $V$ values. The correspondence between the $GW$ and DFT$+U$$+V$(fit) band structures is assessed qualitatively through visual inspection of differences between corresponding bands. The compared features include the alignment of energy levels at critical points in the BZ, the curvature of the bands, and the relative positions of the conduction and valence bands.

\begin{figure}[h!]
\includegraphics[width=\columnwidth]{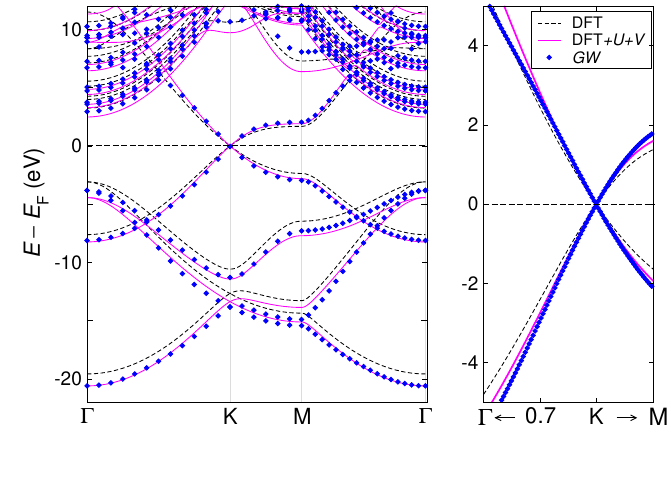}
\caption{\label{fig:fit_Graphene_GW} Band structure of graphene calculated using GGA (dashed black lines) and GGA+$U$$+V$(fit) (magenta lines). The results are compared with $GW$ (blue dots).}
\end{figure}

Using the fit parameters set in Table~\ref{table:table1}, we present $GW$ and DFT$+U$$+V$(fit) band structures in Fig.~\ref{fig:fit_Graphene_GW}. Notably, the valence bands in DFT$+U$$+V$, particularly the $\sigma$-bands, show improved agreement with $GW$. More importantly, at the K-point, the cone-like dispersion is modified to better align with $GW$. This refinement brings the Fermi velocity of the DFT$+U$$+V$(fit) calculation close to the $GW$ Fermi velocity.

\subsection{\label{sec:level5-2}Bulk Si, and Ge}
For Si in the diamond structure a DFT calculation, a $9 \times 9 \times 9$ k-point mesh for BZ sampling, a basis set cut-off of $K_{\text{max}}$=$4.5$~$a_\text{B}^{-1}$, $G_\text{max} = 12.0$~$a_\text{B}^{-1}$, and a MT radius of a Si atom of $2.17~a_\text{B}$) yields good estimates for the lattice constant ($a =5.47$ \AA), and bulk modulus when compared to experimental data, as shown in Table~\ref{table:table3}. The equilibrium lattice constant from DFT is slightly larger than the experimental value of $a =5.431$ \AA, while the bulk modulus is slightly lower, indicating a softer material response compared to experiment. The most notable discrepancy is found between the experimental and DFT-calculated band gaps. To address this point, we first apply DFT$+U$, followed by DFT$+U$$+V$, and evaluate these results in comparison to our $GW$ calculation and other reported findings.

 Employing the equilibrium lattice parameter determined by DFT, we performed cRPA calculations in the subspace of the two $3s$ and $3p$ valence electrons, which form $sp^3$ hybrid orbitals covalently bonded with the hybrid orbitals of the 4 nearest neighbors in a tetrahedral environment, using $u$- and $w$-bases ($G_{\textrm{cut}}=4.5$~$a_\text{B}^{-1}$, and $8 \times 8 \times 8$ k-mesh). Supplementary Table~II~\cite{SM}, and Table~\ref{table:table2} provide the calculated onsite and nearest neighbor intersite bare and partially screened interaction parameters between $s$-$s$, $s$-$p$, and $p$-$p$ orbitals. The results reveal that the values of onsite parameters $U_{ss}$, $U_{sp}$, and $U_{pp}$  are close to each other, as are the intersite parameters $V_{ss}$, $V_{sp}$, and $V_{pp}$, reflecting the hybridized nature of the $s$- and $p$-orbitals. The $V$-values are not small, they are about half as large as the $U$ values in both bases. Again, the $U$-values in $u$-cRPA are larger than the $w$-cRPA values, since the $u$-functions  in the MT are more localized than the Wannier functions. Also, the $V$-values in $u$-cRPA are larger, as their limited spatial extent reduces interatomic screening. Our ($w$ and $u$)-cRPA parameters are notably larger than those reported in Ref.~\cite{campo2010extended}, which were calculated using the linear-response approach. This difference may arise from the distinct treatment of screening and the different projector orbitals in the two methods. In cRPA, screening contributions from the correlated subspace are explicitly excluded, resulting in weaker overall screening and consequently larger interaction parameters. In contrast, the linear-response approach computes interaction parameters using the full electronic susceptibility or dielectric function, incorporating screening contributions from all electronic states. This comprehensive treatment leads to stronger effective screening and thus smaller interaction parameters.

\begin{figure}[h!]
\includegraphics[width=\columnwidth]{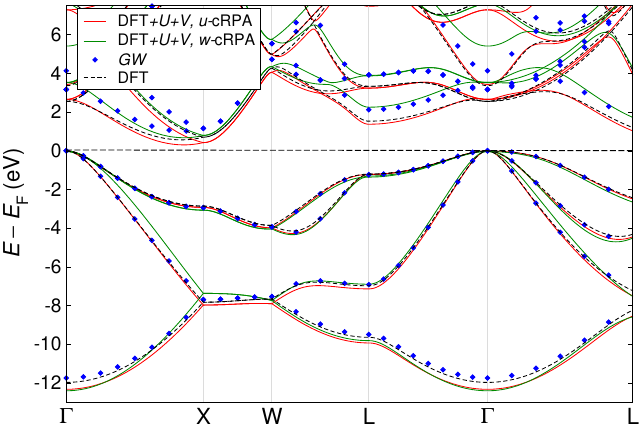}
\caption{\label{fig:Si_dftUV_u_w_GW} Band structure of bulk Si calculated using GGA (dashed black lines) and GGA+$U$$+V$ using $u$- and $w$-cRPA parameters (red and green lines). The results are compared with the $GW$ band structure (blue dots).}
\end{figure}

\begin{figure}[h!]
\includegraphics[width=\columnwidth]{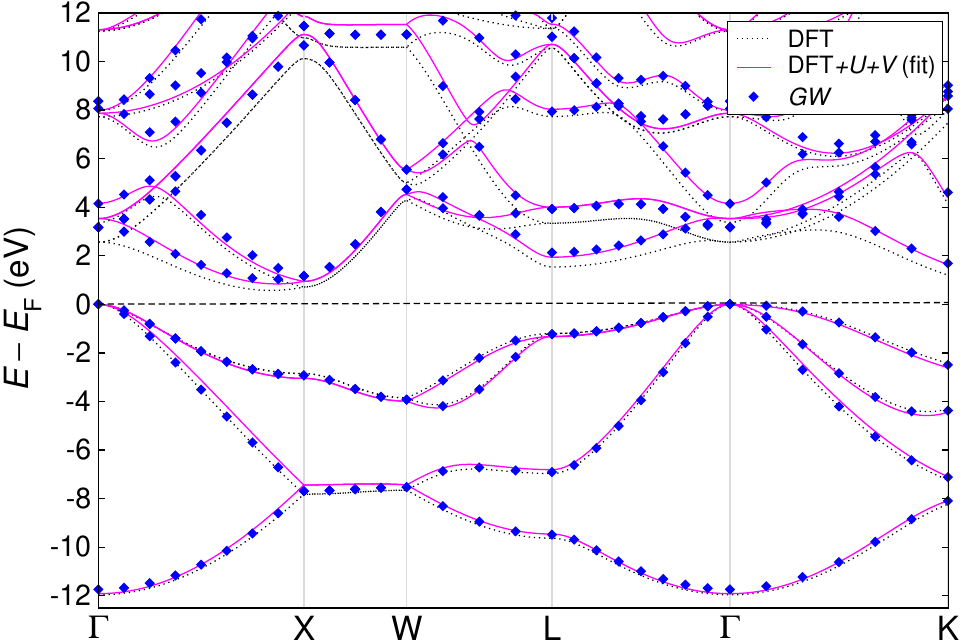}
\caption{\label{fig:fit_Si_gw} Band structure of bulk Si calculated using GGA and GGA+$U$$+V$ utilizing fit parameters. The results are compared to $GW$.}
\end{figure}

Applying DFT$+U$ (using $u$- and $w$-cRPA parameters) results in a lattice expansion, a reduced bulk modulus, and a smaller band gap, worsening agreement with the experimental observations. This behavior can be attributed to the spread of the $p$ states far beyond the MT sphere boundary of the computational setup. Because the formulation of the density matrix in Eq.~\eqref{eq:orbital-occupation_onsite} only considers contributions from within the MT sphere, in such a situation the occupied states appear to be less than half filled in this matrix. In the fully localized limit of DFT$+U$ this leads to a repulsive term. It results in a band closure rather than an opening and weakens the Si-Si bonds, causing a lattice expansion. The effect is more pronounced when $u$-cRPA parameters are used due to their larger values compared to $w$-cRPA. In comparison to DFT$+U$, the inclusion of intersite interactions in DFT$+U$$+V$ enhances Si–Si hybridization, strengthening the bond. As a result, the system exhibits a wider band gap, reduced lattice parameters, and a higher bulk modulus. Fig.~\ref{fig:Si_dftUV_u_w_GW} compares the band structure of Si calculated using DFT, DFT$+U$$+V$ (with both $u$- and $w$-cRPA parameters), and $GW$. The agreement of all calculations with $GW$ in both valence and conduction bands is poor. Also, both direct and indirect band gaps are significantly underestimated. To address this discrepancy, we adjusted the parameters to align DFT$+U$$+V$ more closely with $GW$. Using the fit parameters listed in Table~\ref{table:table2}, the lattice relaxation produces values for $a$, $B$, and $E_{\text{g}}$ that are closer to experimental results. Fig.~\ref{fig:fit_Si_gw} compares the band structure calculated using fit parameters with $GW$, showing improved agreement at the valence band and lower part of the conduction band. Also, direct and indirect gaps are opened ($3.50$~eV and $0.84$~eV) in comparison to $GW$ ($3.16$~eV, and $1.02$~eV, respectively). 

The main differences between our approach and Campo \textit{et al.}~\cite{campo2010extended} lie in the choice of the projector Eq.~\eqref{eq:eq7} and the Hubbard parameters employed. To enable a clearer comparison, we also adopted their parameter set within our projector. The results indicate that the differences can not be attributed solely to the choice of parameters.
\begin{table}[ht]
\caption{\label{table:table3}
Comparison of lattice constant $a$, bulk modulus $B$, and band gap $E_{\text{g}}$ of bulk Si using various approaches. Experimental values are also listed for reference.}
\begin{ruledtabular}
\begin{tabular}{lrrr}
Approach & $a$ (\AA) & $B$ (Mbar) & $E_{\text{g}}$ (eV) \\
\midrule
DFT                         & 5.47 & 88.7  & 0.57 \\
DFT+$U$ ($u$-cRPA)          & 5.96 & 42.9  & 0.01 \\
DFT+$U$ ($w$-cRPA)          & 5.84 & 62.1  & 0.34 \\
DFT$+U$$+V$ ($u$-cRPA)        & 5.16 & 75.2  & 0.21 \\
DFT$+U$$+V$ ($w$-cRPA)        & 5.02 & 102.7 & 0.68 \\
DFT$+U$$+V$ (fit)             & 5.39 & 92.2  & 0.84 \\
DFT$+U$$+V$~\cite{campo2010extended} & 5.37 & 102.5 & 1.36 \\
Experiment~\cite{ioffe_semiconductors} & 5.43 & 98.0  & 1.12 \\
\end{tabular}
\end{ruledtabular}
\end{table}
\begin{table}[ht]
\caption{\label{table:table2}
Partially screened onsite and first-nearest-neighbor intersite interaction parameters for bulk Si and Ge, calculated using cRPA with $u$- and $w$-functions. Si values are compared with those from Refs.~\cite{campo2010extended, tancogne2020parameter}. Interaction indices denote interactions within or between $s$- and $p$-orbitals. The sixth and ninth columns list the fitted parameters used to match $GW$ results as described in the text.}
\begin{ruledtabular}
\begin{tabular}{lrccccrcc}
(eV) & \multicolumn{5}{c}{Si} & \multicolumn{3}{c}{Ge} \\
\cline{2-6} \cline{7-9}
 & $u$-cRPA & $w$-cRPA & ~\cite{campo2010extended} & ~\cite{tancogne2020parameter} & fit 
     & $u$-cRPA & $w$-cRPA & fit \\
\midrule
$U_{ss}$ & 10.66 & 8.51 & 2.82 & 3.68 & 3.78 & 10.36 & 8.57 & 4.15 \\
$U_{pp}$ & 10.34 & 6.74 & 3.65 & 3.55 & 3.66 & 9.92  & 6.23 & 3.66 \\
$U_{sp}$ & 10.48 & 7.45 & 3.18 & 2.29 & 2.00 & 10.11 & 7.13 & 3.83 \\
$V_{ss}$ & 4.89  & 4.74 & 1.40 & 0.94 & 1.40 & 4.65  & 4.52 & 4.00 \\
$V_{pp}$ & 4.88  & 4.38 & 1.34 & 1.86 & 3.00 & 4.64  & 4.05 & 4.00 \\
$V_{sp}$ & 4.89  & 4.55 & 1.36 & 1.37 & 1.30 & 4.64  & 4.27 & 0.68 \\
\end{tabular}
\end{ruledtabular}
\end{table}

\begin{figure}[h!]
\includegraphics[width=\columnwidth]{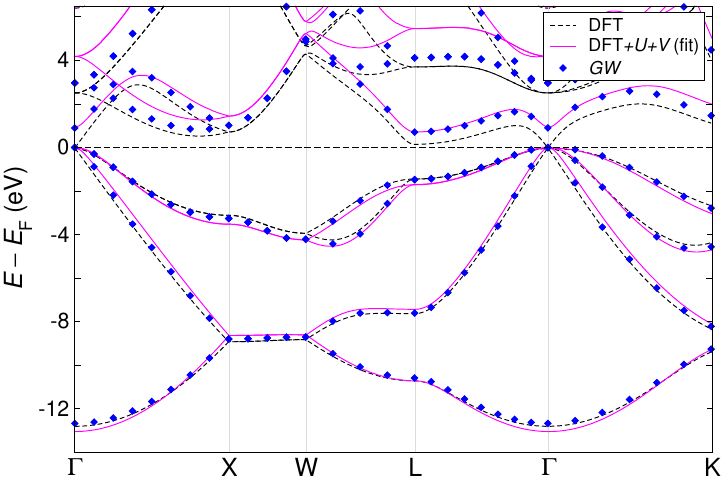}
\caption{\label{fig:fit_Ge_gw} Band structure of bulk Ge calculated using GGA and GGA+$U$$+V$ utilizing the fitted parameters. The results are compared to the $GW$ band structure.}
\end{figure}

A scenario similar to bulk Si is observed when analyzing bulk Ge in Table~\ref{table:table5} (diamond structure, lattice constant $a=5.65$ \AA, MT radius of Ge atom of $2.25~a_\text{B}$) using the same type of parameters as in bulk Si. Table~\ref{table:table2} presents the cRPA parameters for Ge, which were subsequently used to investigate their impact on $a$, $B$, and $E_{\text{g}}$. Both DFT$+U$ and DFT$+U$$+V$, based on the two parameter sets, exhibit similar trends as studied in Si-case. Notably, the inclusion of non-local interactions leads to an opening of the band gap, although the resulting value still deviates significantly from experimental and $GW$ results. However, by employing a parameter set optimized to match the $GW$ band structure (specially the occupied bands and lowest conduction band), we achieved improved estimates for $B$ and $E_{\text{g}}$ that align more closely with experimental data. Fig.~\ref{fig:fit_Ge_gw} compares the $GW$ with DFT$+U$$+V$ (using fit parameters) band structures. The inclusion of SOC into the last calculation results in a band gap of 0.58~eV. 

It is important to highlight that for the fit parameters of Si and Ge, we required $V$ values comparable to $U$, although on-site parameters should generally be larger than intersite parameters.
The fact that $V\approx U$ points to the possibility that nearest-neighbor interactions are not sufficient and next-nearest-neighbor interactions would have to be included. Indeed, given the spatial extent of the $sp^3$ hybridized orbitals, this is perhaps not surprising. 

\begin{table}[ht]
\caption{\label{table:table5} Comparison between lattice parameters, $a$, bulk modulus, $B$, and band gap, $E_{\text{g}}$, of bulk Ge using different approaches.}
\begin{ruledtabular}
\begin{tabular}{lccc}
 Approach & $a$(\AA) & $B$(MBar)  & $E_{\text{g}}$ (eV)\\ 
 \midrule
DFT              & 5.746 & 71.9 & 0.000\\
DFT$+U$ $(u$-cRPA)  & 6.379 & 39.6 & 0.000 \\
DFT$+U$ $(w$-cRPA)   & 6.186 & 36.6 & 0.001\\
DFT$+U$$+V$ $(u$-cRPA) & 5.588 & 18.8 & 0.002\\
DFT$+U$$+V$ $(w$-cRPA)  & 5.308 & 84.0 & 0.571 \\
DFT$+U$$+V$ (fit)  & 5.364 & 79.9 & 0.713 \\
Experiment.~\cite{jivani2005some},~\cite{yuan2024direct}             & 5.658 & 75.0 & 0.800\\
\end{tabular}
\end{ruledtabular}
\end{table}

\subsection{\label{sec:level5-3}NiO}
We now turn our attention to a transition-metal oxide NiO, the prototype Mott-charge-transfer insulator. NiO crystallizes in the cubic rock-salt structure. Our calculations were performed assuming that the magnetic moments of the Ni atoms adopt a type 2 antiferromagnetic (AFM-2) arrangement, which is observed experimentally below the Néel temperature ($T_{\text{N}}=523$ K). This configuration is characterized by alternating ferromagnetic planes stacked along the [111] direction. We used a $16 \times 16 \times 16$ k-mesh. 
For the basis of the KS orbitals, the plane-wave cutoff  was set to $K_\text{max}=4.3~a_\text{B}^{-1}$, $G_\text{max} = 12.5$~$a_\text{B}^{-1}$, with angular-momentum expansions inside the muffin-tin spheres truncated at $\ell_\text{max}=10$ for Ni and $\ell_\text{max}=6$ for O. The muffin-tin radii were chosen as $R_\text{MT}^\text{Ni}=2.32~a_\text{B}$ and $R_\text{MT}^\text{O}=1.32~a_\text{B}$. The density of states (DOS) was calculated using the tetrahedron method for Brillouin-zone (BZ) integration.

\begin{figure}[h!]
\includegraphics[width=\columnwidth]{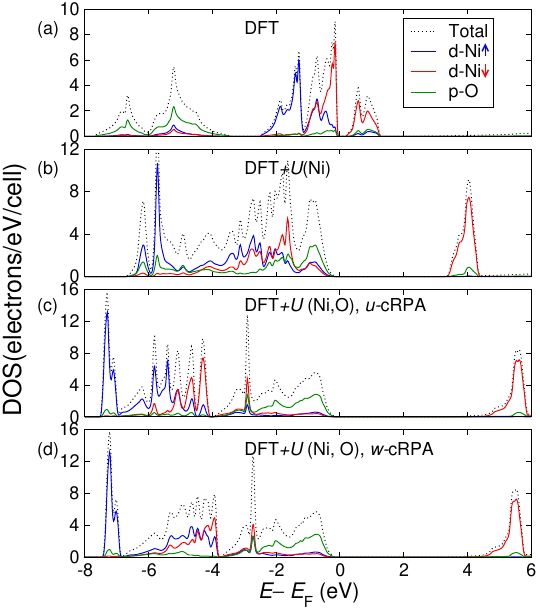}
\caption{\label{fig:DFT_U_dos}DOS of NiO obtained with different approaches: (a) DFT, (b) DFT+$U$(Ni) using $U$ from $w$-cRPA, (c) DFT+$U$(Ni)+$U$(O) using $u$-cRPA parameters, (d) DFT+$U$(Ni)+$U$(O) using $w$-cRPA parameters. All energies are relative to the Fermi level.}
\end{figure}

At the DFT level, the equilibrium lattice constant shows a slight deviation from the experimental value, as indicated in Table~\ref{table:table7}. The bulk modulus falls within the accepted experimental range. In contrast, DFT significantly underestimates the band gap and the magnetic moment of the Ni atoms. Given the charge-transfer nature of NiO, the top of the valence band is expected to be primarily composed of O $p$-orbitals, while the conduction band should be dominated by Ni $d$-orbitals. This would allow for electron excitation from $p$-O to $d$-Ni across the band gap~\cite{rohrbach2004molecular}. However, this expected spectroscopic behavior is not reflected in the DOS of DFT as shown in Fig.~\ref{fig:DFT_U_dos}(a), where the spin-down Ni states dominate the edges of both the valence and conduction bands. Additionally, the peak at $-5$~eV relative to the Fermi level shows the presence of $p$-orbitals of O, which contradicts the experimentally observed spectroscopic features dominated by $t_{2\text{g}}$ states of Ni~\cite{fujimori1984valence, sawatzky1984magnitude}.

Table~\ref{table:table6} presents the Hubbard parameters calculated using cRPA in both bases. In the $u$-basis, orbitals are confined within the atomic spheres. Again, this localization leads to stronger interaction parameters. Conversely, the Wannier basis accounts for the hybridization between Ni and O orbitals and their delocalization leading to reduced $U$ and $V$ values. To directly compare the implications of this model with previous studies where $U$ was applied solely to Ni, we performed cRPA calculations in $w/u$ bases while including only the $d$-orbitals of Ni. In this restricted model, $U$(Ni) was found to be $5.5/6.5$~eV for both bases respectively.
\begin{table}[ht]
\caption{\label{table:table7} Comparison between lattice parameters, $a$, bulk modulus, $B$, fundamental band gap, $E_{\text{g}}$, and magnetic moment of Ni atoms, $m_{\mathrm{Ni}}$, in NiO (AFM-2) using different approaches.
a: Ref.~\cite{lide2004crc}, b: Ref.~\cite{gavriliuk2023first}, ~\cite{sawatzky1984magnitude}, c: Ref.~\cite{sawatzky1984magnitude},~\cite{kurmaev2008oxygen}, d:~\cite{cheetham1983magnetic,uno1963structure} }
\begin{ruledtabular}
\begin{tabular}{lcccc}
 Approach & $a$ & $B$  & $E_{\text{g}}$ & $m_{\mathrm{Ni}}$ \\ 
 & (\AA) & (MBar) & (eV) & ($\mu_{\mathrm{B}}$)\\
\midrule
DFT                 & 4.19 & 200 & 1.35 & 1.37 \\
DFT$+U$ ($u$-cRPA)  & 4.23 & 167 & 3.09 & 1.81 \\
DFT$+U$ ($w$-cRPA)  & 4.23 & 170 & 3.07 & 1.80 \\
DFT$+U$ (Ni)        & 4.21 & 179 & 3.05 &  1.70 \\
DFT$+U$$+V$ ($u$-cRPA)& 4.19 & 172 & 3.34 & 1.80 \\
DFT$+U$$+V$ ($w$-cRPA)& 4.17 & 181 & 3.53 & 1.78 \\
DFT$+U$$+V$~\cite{campo2010extended}  & 4.25 & 189 & 3.6 & - \\
DFT$+U$$+V$~\cite{campo2010extended} (SC) & 4.23 & 197 & 3.2 & - \\
Experiment        & 4.18$^{\text{a}}$ & 166-208$^{\text{b}}$ & 4.0,4.3$^{\text{c}}$ & 1.64,1.90$^{\text{d}}$\\
\end{tabular}
\end{ruledtabular}
\end{table}

\begin{table}[h]
\caption{\label{table:table6} Partially screened onsite and intersite interaction parameters for NiO, computed using cRPA in $u$- and $w$- bases. The calculations include the $d$-orbitals of Ni and the $p$-orbitals of O. Results are compared with previously reported values from Ref.~\cite{campo2010extended}.} 
\begin{ruledtabular}
\begin{tabular}{ccccc}
Parameter (eV) & $u$-cRPA & $w$-cRPA& Ref.~\cite{campo2010extended} \\
\midrule
$U$ (Ni)  & 9.22 & 9.02 & 7.27  \\ 
$U$ (O)   & 8.45 & 5.91 & -  \\ 
$V$ (Ni-Ni)  & 1.50 & 1.50 & 0.77 \\ 
$V$ (Ni-O)   & 1.21 & 2.23 & 1.91  \\ 
$V$ (O-O)    & 1.51 & 1.51 & -  \\ 
\end{tabular}
\end{ruledtabular}
\end{table}

Incorporating a $U$ of $5.5$~eV for Ni atoms (line 4 in Table.~\ref{table:table7}) leads to several notable effects in comparison with DFT: a lattice expansion (exceeding the experimental value), a bulk modulus that is softer, a significant band gap opening to $3.05$~eV, and a significant increase of the magnetic moment for Ni atoms. Examination of the DOS in Fig.~\ref{fig:DFT_U_dos}(b) reveals an increased presence of O $p$-orbitals at the top of the valence band compared to standard DFT, however, there is no satellite at $-7$~eV as observed experimentally~\cite{thuler1983photoemission}. Applying a $U$ value of $6.5$~eV amplifies these effects, it does not bring the results into closer alignment with experimental observations. Extending the treatment by applying a $U$ to both Ni and O orbitals using cRPA parameters from Table~\ref{table:table6} further enhances these effects, yielding results consistent with those observed in DFT+$U$(Ni) for the lattice parameter, bulk modulus, band gap, and magnetic moment. Notably, $p$-states of O dominate the top of the valence bands, aligning with experimental findings~\cite{sawatzky1984magnitude}. The mixed character at the top of the valence band is confirmed by X-ray photoelectron spectroscopy (XPS) experiment~\cite{van1992electronic}. The peak due to occupied t$_{2\text{g}}$ below $-2.5$~eV agrees with the study of NiO within the DFT+DMFT approach~\cite{ren2006lda+}. Furthermore, the position of a pronounced peak (with $e_{\text{g}}$ character) at approximately $-7$~eV coincides with an observed satellite in~\cite{thuler1983photoemission}.
\\

\begin{figure}[h!]
\includegraphics[width=\columnwidth]{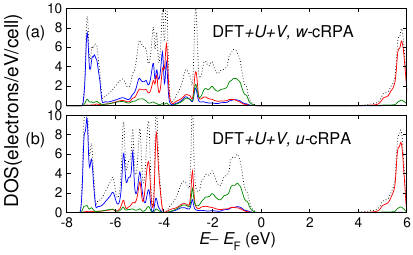}
\caption{\label{fig:DFT_UV_dos} NiO DOS obtained using: (a) DFT$+U$$+V$ using $w$-cRPA parameters; (b) DFT$+U$$+V$ using $u$-cRPA parameters. All energies are relative to the Fermi level.}
\end{figure}

Our DFT$+U$$+V$ calculations, using cRPA-derived parameters in both bases, successfully address the limitations of both DFT and DFT$+U$. Compared to DFT$+U$, the inclusion of the inter-site interaction $V$ leads to a reduced lattice parameter, bringing it closer to the experimental value and in agreement with the result reported in~\cite{campo2010extended}. Additionally, the band gap is further widened relative to DFT($+U$). This system has been studied using various hybrid functionals, as PBE0, HSE03, HSE06, MHSE, DD-RSH-CAM, and DSH0. They yield fundamental band gaps of $5.29, 4.28, 4.56, 5.72, 4.34,$ and $4.16$~eV, respectively~\cite{liu2019assessing}. The band gap improvement of  hybrid functionals arises from their incorporation of non-local exact exchange, which better captures the electron interactions and corrects the self-interaction error. However, certain hybrid functionals tend to overestimate the band gap. Referring to Fig.~\ref{fig:DFT_UV_dos}, all the features enhanced in alignment with experimental observations in DFT$+U$(Ni,O) are also present in DFT$+U$$+V$ (both bases). The increased bandwidth in the DFT$+U$$+V$ DOS indicates a stronger hybridization between the $p$- and $d$-orbitals. 
\\
\\

\begin{figure}[h!]
\includegraphics[width=\columnwidth]{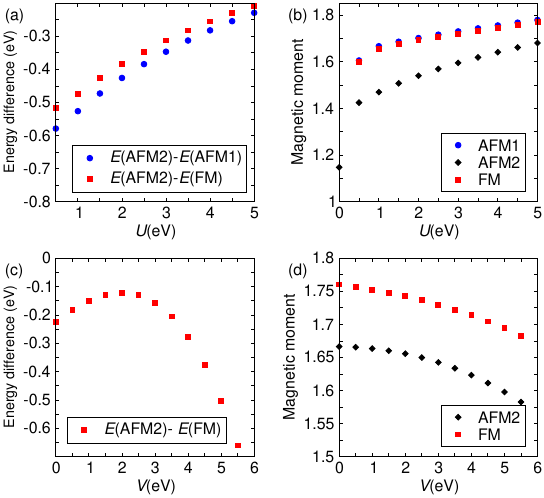}
\caption{\label{fig:change_UV} (a) The energy difference between AFM-2 and AFM-1 phases, and between AFM-2 and FM phases of NiO, as a function of the onsite $U$ on the Ni atoms. (b) The magnetic moments of Ni atoms ($\mu_{\mathrm{B}}$) in AFM-2, AFM-1, and FM phases as a function of $U$.
(c) The energy difference between AFM-2 and FM phases as a function of the intersite $V$ between Ni and O atoms. (d) The magnetic moments of Ni atoms ($\mu_{\mathrm{B}}$) in AFM-2 and FM phases as a function of $V$. All calculations were performed using the same unit cell with fixed lattice parameters to ensure a consistent comparison between magnetic phases.}
\end{figure}

NiO, beyond the studied AFM-2 phase, can also exhibit alternative magnetic configurations, including the FM phase and AFM-1 phase. In the latter, ferromagnetic planes are stacked antiferromagnetically along the [001] direction. We aim to explore the role of Hubbard corrections on phase stability, and magnetic properties. To this end, as depicted in Fig.~\ref{fig:change_UV} (a-b), we vary $U$ in the three magnetic configurations. Up to the studied range of $U$, AFM-2 is the preferred phase. The relative stabilization of the AFM-2 phase over the other phases weakens as $U$ is increased. The magnetic moments of the Ni atoms interact indirectly through the oxygen anions in the lattice via super-exchange. AFM-2 is energetically favored over AFM-1 due to the geometry of the Ni fcc-lattice, which supports this magnetic ordering, and the positioning of the oxygen atoms, which facilitates superexchange interactions. The magnetic moments for all phases increase steadily with $U$ indicating enhanced localization of electrons. Using $U = 4.6$~eV for Ni atoms, we systematically varied the value of $V$ between the Ni and O atoms at the nearest-neighbor level for both AFM-2 and FM configurations. Across all values of $V$, the AFM-2 phase remains the most energetically stable. The increase of $V$ strengthens the superexchange interaction, further stabilizing the AFM-2 phase. At the same time, the magnetic moments in both AFM-2 and FM configurations decrease as $V$ increases. This reduction is attributed to the enhanced hybridization, which facilitates the delocalization of electrons in the Ni $d$-orbitals, leading to a decrease in the magnetic moment.
\section{\label{sec:level6}Conclusion}

Motivated to overcome the limitations of DFT in describing strongly correlated electronic systems, systems with charge transfer and charge disproportionation, we extended the DFT$+U$ framework by implementing a DFT$+U$$+V$ functional within the all-electron FLAPW method, as realized in the FLEUR code. The interaction parameters are determined from first principles using  the constrained random phase approximation (cRPA) concept as implemented in the FLAPW code SPEX. Two  different functions are employed to span the correlated subspace: the  muffin-tin $u$-functions (MTF) or the maximally localized Wannier functions (MLWF). The $u$-function basis ensures consistency with the LAPW basis set, whereas the Wannier basis establishes a more direct link to low-energy model Hamiltonians. A potential drawback of the $u$-function representation is its dependence on the muffin-tin radius, which causes the resulting $U$ and $V$  parameters to vary with the chosen radius. We find that, upon increasing the muffin-tin radii, the $u$-cRPA interaction parameters converge toward those obtained with the $w$-cRPA approach, in line with trends reported for pseudopotential-based implementations. In this work, the subspace projection is restricted to spin-diagonal terms. Our implementation is therefore applicable to systems in which spin–orbit coupling and non-collinear magnetism can be safely neglected.

The methodology was comprehensively validated for graphene, bulk Si, Ge, and NiO, representing two- and three-dimensional materials, covalently and ionically bonded, non-magnetic and magnetic materials.  We compared the DFT$+U$ and the DFT$+U$$+V$ model for the MTF and the MLWF representation of the correlated subspace with DFT, $GW$, experimental results as well as results of other computational frameworks, when ever possible, allowing for a direct comparison. In general, we find good agreement with previous results, and we find a small dependence of the calculated physical properties on the two representations. 

2D material: In the case of graphene, the DFT$+U$$+V$ correction refines the slope of the Dirac cone, resulting in a Fermi velocity closer to the experimental value, which is a critical factor for transport properties. The refinement has a slight dependence on the representation of the correlated space, but considering the difference in the $U$ and $V$ parameters in the two different representations, the difference is really small. 

Elemental semiconductors: For bulk semiconductors like Si and Ge, the approach demonstrated its ability to handle systems with significant covalency  of $sp$ bonds. DFT gives a close estimate to the experimental lattice parameter, but underestimates the gap, while the $+U$ correction open the gap, but at the same time it is contra productive  for the accuracy of the lattice constant. The $+V$ correction improves the lattice constants and electronic band gaps, bringing them into better agreement with experiment. 

Magnetic insulator: In NiO, the intersite $V$ term captures besides the Coulomb repulsion between the Ni $d$-electrons controlled by $U$, the charge transfer energy between the O $2p$- and the Ni $3d$-states and leads to a more accurate description of the splitting between occupied and unoccupied $d$-states. These findings validate the utility of the DFT$+U$$+V$ framework in systems where electron localization and hybridization both play crucial roles. In NiO, DFT offers structural accuracy, while DFT$+U$ improves electronic properties, and magnetic moments at the expense of structural precision.  The DFT$+U$$+V$ framework overcomes the limitations of both approaches simultaneously—improving lattice constants,  magnetic moments and band gaps of NiO, and bringing them into closer agreement with experimental values.  Overall, across the range of $U$ and $V$ values considered, no phase transition occurs, and the AFM-2 phase remains the ground state.

The DFT$+U$$+V$ approach serves a good compromise between improved accuracy of many important physical quantity including the band dispersion and the band gap, and computational efficiency, allowing a scaling to complex materials.
In future work, we aim to explore self-consistent evaluation of Hubbard parameters, further enhancing the predictive power and internal consistency of the method. Depending on the requirements on improving the band-dispersion over a large energy width, lifting the static limit in the definition of the screened Coulomb interaction might a future direction.

\begin{acknowledgments}
We gratefully acknowledge fruitful discussions with Ersoy \c{S}a\c{s}{\i}o\u{g}lu, Hongbin Zhang, and Motoaki Hirayama. C.F. gratefully acknowledges computing time on the supercomputer JURECA~\cite{JURECA} at Forschungszentrum Jülich under grant Topomag. This work was supported by several funding sources, including the Deutsche Forschungsgemeinschaft (DFG) through CRC 1238 (Project C01), the Federal Ministry of Education and Research of Germany (BMBF) within the framework of the Palestinian-German Science Bridge (BMBF Grant No. DBP01436), and the European Centre of Excellence MaX (Grant No. 101093374).
\end{acknowledgments}

%
\end{document}